\documentclass{svjour3}                   
\smartqed  %
\usepackage{epsfig}
\usepackage{graphicx}
\usepackage{lineno}
\usepackage{amsmath}
\usepackage{url}
\usepackage{color}
  
\newcommand{\black}{\color{black}{ }} 
\newcommand{\blue}{\color{blue}{ }} 
\usepackage[a4paper,width=150mm,top=25mm, bottom=25mm]{geometry}
\journalname{Journal:}

\begin{document}

\title{Policy lessons from the Italian pandemic of Covid-19}
\author{Jos\'{e} M. Carcione$^{1,2}$ \and Jing Ba$^{1(*)}$}
\institute{
$^1$Hohai University, Nanjing, China \\
$^2$Associated with the National Institute of Oceanography and Applied Geophysics - OGS, Trieste, Italy. \\
(*) Corresponding author. Email address: jba@hhu.edu.cn}

\date{Submitted: \today \ }

\maketitle 

\baselineskip 15pt

\parindent 0.in

{\bf Running title}: Management of the COVID-19 pandemic.

\vspace{0.5cm}


\vspace{1cm}


\begin{abstract}
We analyze the management of the Italian pandemic during the five identified waves. We considered the following problems: (i) The composition of the CTS (``Scientific Technical Committee"), which was composed entirely of doctors, mainly virologists, without mathematical epidemiologists, statisticians, physicists, etc. In fact, a pandemic has a behavior described by mathematical, stochastic and probabilistic criteria; (ii) Political interference in security measures and media propaganda; (iii) The initial stages of the vaccination campaign, ignoring the age factor, and (iv) The persistence of the pandemic due to the population unvaccinated (anti-vax or ``no-vax"), which amounts to about six to seven million people, including 10\% of anti-vax doctors.
\end{abstract}

\section{Introduction}    

The World Health Organization (WHO) declared the novel coronavirus disease 2019 (COVID-19) a Public Health Emergency of International Concern on January 30, 2020 and a pandemic on March 11, 2020. The pandemic began in late 2019 in Wuhan (China), where the severe acute respiratory syndrome Coronavirus 2 was first identified. By the end of September 2022 the WHO reported 616,066,645 detected cases and 6,531,468 deaths globally (Cucinotta and Vanelli, 2020). 

The pandemic started in the West in northern Italy (Lombardy region), on February 24, 2020, with a strong contagion rate. Right after the first wave, we implemented
a SEIR epidemiological model to simulate the pandemic in Lombardy and calculate the infected population and the mortality rate (Carcione et al., 2020). The model was calibrated based on the total number of casualties. The results indicate an infection fatality rate of 0.00144/day (IFR = 0.57\%).
Subsequent work has confirmed this finding (e.g., O'Driscoll et al., 2021). These studies reveal that the death rate of COVID-19 can be up to seven times that of the flu.

Despite declaring a national health emergency on January 31, 2020, first official public data date back to February 24, 2020 and by September 30, 2022, health institutions recorded 22,605,873 confirmed cases and 173,430 deaths distributed on five waves$\footnote{\url{https://github.com/floatingpurr/covid-19_sorveglianza_integrata_italia/blob/main/data/2022-09-30/data.xlsx}}$
Health policies evolved from a national lockdown to a population vaccination campaign accompanied by local restriction measures (according to territorial epidemiological measures). The national health emergency ended on March 31, 2022. As of September 30, 2022, the only COVID-19 preventive measure maintained is the requirement to wear surgical masks in hospitals and nursing homes (all the others were gradually removed).

Around the world, misinformation was rife, and questionable policies were sometimes enforced at national and subnational level. For example in the United States (Kapucu and Moynihan, 2021) and in Brazil, where politics played a more negative role than in Italy, since they undermined the credibility of the federal government. Inconsistencies between states have meant different access to personal protective equipment or coordination of vaccine distributions.

This paper aims to evaluate Italy's pandemic management during the first five waves of deaths using current knowledge and suggest improvements for post-COVID-19 pandemic planning.
We consider various problems, namely the composition of the CTS, the political interference in the safety measures, the lack of adequate information, the vaccination campaign and the persistence, still today, of the pandemic due to the unvaccinated (``anti-vax") . 

\section{Policy options in Italy and implications}

In this section, we discuss the main issues in Italian health policies that could teach us important lessons for the future. Our concerns include: the evaluation of the delay in seeing the effects of health policies; the composition of the CTS; political interference in safety measures; a lack of sufficient scientific information; the vaccination campaign; and dealing with the anti-vaccination attitude.

Mathematical modeling is essential because it can be used to predict different scenarios (of deaths, hospitalizations, infections) based on different assumptions (hard lockdown, soft lockdown, social distancing). The equations governing a pandemic are of a diffusive nature, widely used in many fields of physics, such as geophysics, etc. (e.g. Carcione, 2022). 

Carcione et al. (2020) considered a SEIR epidemic disease model, where
reported infected people were not used for calibration, as this data was unreliable, and by then
the number of undiagnosed asymptomatic infections was largely unknown. In fact, we used the number of casualties for calibration, since it is more robust.
Hospitalizations are related to the number of victims, but it is better to use the latter for calibration, since death can occur before hospitalization (as in China during the first wave), i.e. people who never reach the hospital. There are models that use hospitalizations and deaths for calibration (e.g., Morozova et al., 2021). 

As shown in Carcione et al. (2020), the pandemic develops in a few days and the death toll can be extremely high if the mortality rate and contagiousness of the disease are high. Just a few days before taking action can make all the difference in preventing this disaster. The pandemic and its consequences were predicted in October 2019 by a group of experts (Brannen and Hicks, 2020). However, states have ignored this fact and the transnational nature of the threat, delaying the necessary measures. In less than three weeks, the virus has overwhelmed the healthcare system in northern Italy, especially in Lombardy.

\subsection{Scientific and Technical Committee}

On 5 February 2020, the CTS was established to support the central government in activities aimed at overcoming the epidemiological emergency due to the spread of the Coronavirus.
The members include doctors, lawyers, mainly executives, but not experts in mathematical epidemiology$\footnote{\url{https://www.salute.gov.it/portale/nuovocoronavirus/dettaglioContenutiNuovoCoronavirus.jsp?area=nuovoCoronavirus&id=5432&lingua=italiano&menu=vuoto}}$

Mathematicians, software designers and physicists (mathematical epidemiologists) were needed to analyze the transmission rates and distribution patterns of the virus.
According to Pistoi (2021), the CTS lacked experience in other crucial areas, apart from epidemiology.
SAGE (Scientific Advisory Group for Emergencies) from the UK comprises specialists in molecular diagnostics, high-throughput sensing, sequencing, modeling, logistics,
behavioral and educational sciences. The French government advisory board includes a digital technology specialist,
anthropologists and sociologists together with basic skills in virology, epidemiology and molecular biology. The Accademia Leopoldina, a top adviser to the German government, has several working groups that chosen from a list of over 1600 international members in any discipline.

Furthermore, as we will see later, the vaccination campaign launched by the CTS in the first four months was based on the criterion of contagion and not on the risk of death.

\subsection{The five main waves}

Several miscalculations were due to the lack of an adequate pandemic plan. The issues related to the five main waves (see Figure 1), indicating the date of the peak of deaths, are:

\begin{enumerate}

\item
First wave (March 31, 2020): The total lockdown began on March 22 and should have started at least a month earlier, given the experience of Wuhan (China), to avoid thousands of deaths. In China, the lockdown began on January 23, 2020, with the peak of deaths on February 12.

\item
Second wave (November 23, 2020): It was not expected and it was worse than the first. The (partial) lockdown began on 24 October 2020, about a month late.

\item
Third wave (April 13, 2021): Inadequate vaccination campaign (see below).

\item
Fourth wave (February 4, 2022): Compulsory vaccination (over 60/65) delayed by one year.

\item
Fifth wave (July 29, 2022): from May 1, the green pass (vaccination certificate) is no longer required to access services and activities, increasing the infected population.

\end{enumerate}

\subsection{Contact-tracing}

By following the WHO guidelines, initial testing strategy focused on suspected cases linked with China, thus limiting the country's diagnostic capacity in the most effective period of action. The "Immuni" application did not work for several reasons, one of which was the alleged violation of privacy and inappropriate media propaganda, which led the majority of Italians not to download the application. (Isonne et al., 2022). The daily public data of the epidemiological surveillance system (deaths, hospitalizations and infections) do not contain details on gender and age and only the quartiles of distributions of times from infection to diagnosis, symptoms, hospitalization and deaths were provided. This limited the potential contribution from the scientific community.

\subsection{The vaccination campaign} 

The risk of death increases approximately as a Gaussian curve from age 60-65 (see Figure 2). 
However, at the beginning, the CTS  recommended getting vaccinated due to the risk of contagion (the so-called categories: school employees, lawyers, etc.) and not due to the risk of death.
In fact, the CTS recommended giving AstraZeneca to young people between the ages of 18 and 55 on February 1, 2020. And teachers and other public workers have also been vaccinated since March 2020 in parallel with the over 80s$\footnote{\url{https://www.aifa.gov.it/documents/20142/1289678/Vaccino-AstraZeneca_parere_CTS\30.01-01.02.2021.pdf}}$. 
As of May 2021, about 10 million people had been vaccinated, half of whom were under the age of 65$\footnote{\url{https://github.com/pcm-dpc/COVID-19}}$ (see Figure 3). It is clear that there was no age priority and that the vaccination was carried out on the basis of the risk of contagion.
Italy could have fully vaccinated its elderly population (65$+$ years) in two to three months if it had vaccinated half a million people per day from January 2021. 
Certainly, vaccine scarcity and government crisis (from January 13th to February 13th in 2021)  hindered the rollout of the vaccination campaign (Figure 3). Defining collaboration with the EU Commission to secure a steady vaccine supply at European level and regional implementation in the national pandemic plan could reduce those adverse risks.

Italian mathematicians (Faranda et al., 2021) calculated the additional deaths resulting from missed vaccinations using the SEIR model. The conclusions are that the excess of deaths is about 2500 if they are vaccinated with a delay of 30 days.
It would have been enough to vaccinate all the over 65s and the pandemic would have ended.
In the best case scenario, if from January 2021 we had proceeded quickly with half a million vaccinated a day, i.e. 15 million in 30 days, the group of people over 65 could have been vaccinated in a month, and  thousands of victims could be avoided. In February and March 2020 alone, 15,000 deaths were recorded, an average of 250 deaths per day.

The defamation of the vaccine (such as serum, drugs, microchips, etc.) by the media (mainly social media) and some politicians is exemplified by the AstraZeneca case. There were 7 deaths from thrombosis in 18 million people vaccinated with this vaccine, 
compared to 1.6 per 1000 annually (Waheed et al, 2022). 
It is difficult to prove the cause and effect link. However, if that were the case, the chance of dying would be 1 in 2.5 million, while the chance (in 80 years) of being hit by a meteor would be 1 in 1.6 million; in a traffic accident, 1 in 90; in a fire, 1 in 250; from a tornado, 1 in 600,000, or from a lightning strike, 1 in 100,000.
Failure to vaccinate with the AstraZeneca vaccine in Italy has caused thousands of deaths (e.g., Faranda, et al., 2021).
Was the cause a drug war between companies?

\subsection{The ``anti-vax" problem} 

Figure 4 shows the number of people who died in a 7-day average around December 6, 2020 (no vaccine, black squares and line) and around December 6, 2021 (vaccination present, dots and red line) as a function of the total population ( in millions of people). Each symbol corresponds to a European country. They are 13: Italy, France, United Kingdom, Germany, Poland, Austria, Czech Republic, Spain, Portugal, Switzerland, Netherlands, Sweden and Belgium. The lines correspond to a linear fit of the data (squares and points). Among the black squares we exclude the UK (outlier) with 1200 deaths, as Prime Minister Boris Johnson in 2020 called for ``freedom for all", clearly departing from standard measures. Among the red dots, we exclude Poland (another outlier), on December 6 with almost 400 deaths for having vaccinated less than 50\% of the population.
It is clear that, with the vaccine, deaths have decreased significantly, which shows that this is the solution to the problem. For Italy (60 million), the red line indicates around 100 victims per day.
These deaths correspond mostly to the anti-vaccine population, a small country within a large one.
In Italy in September 2021 there were nearly 12\% million against vaccination$\footnote{\url{https://www.ispionline.it/it/pubblicazione/datavirus-quanti-sono-i-no-vax-europa-31530}}$, 
of which about 2-3 million were over 65 years old, with a high risk of death. The situation was no different recently, on October 30, 2022, with 72 victims. Paradoxically, the green pass protects those who are against it, i.e. anti-vaccination, because it prevents them from being in crowded places, and indirectly protects the entire population because it prevents the increase of Covid patients in the intensive care units.

\subsection{Examples of typical media disinformation.} 

In some television ``talk-shows'' it has been claimed that 67\% of hospitalized patients (in intensive care and deceased) are anti-vax and 33\% are vaccinated, which would demonstrate that the vaccine does not work in 1 person out of 3. However, it must be taken into account that 10\% of the population was anti-vax compared to 90\% of the vaccinated, therefore the virus affects 54 million vaccinated and 6 million anti-vaccinated. Then, the proportion of 1 to 3 must be weighed with a factor of 90\% / 10\% = 9 and the true fact is that the vaccine does not work in 1 person out of 27. 
In addition, there was the false claim that the vaccine is ineffective because it does not prevent 100\% of infections, ignoring its ability to significantly reduce the risk of severe disease. 
Again, the issue of contagion instead of considering the substantial reduction in mortality risk.
We can see how misinformation and media propaganda have induced an unfavorable situation to deal with pandemics to make people avoid the vaccine.

An important factor in the successful management of a pandemic is the public availability and quality (of presentation) of epidemiological data to the general public. In Italy, doctors and virologists have been the main protagonists of the media, which have constituted a complexity of voices that has added much confusion to the understanding of the crisis. Also, 
authoritative personalities and institutions gave insufficient,
often conflicting information and advice that were later
reversed. On rare occasions, epidemiologists might present their case. For example, a graph like the one in Figure 2 was hardly shown and explained, if ever, that it clearly would convince the authorities to vaccinate the elderly as soon as possible and that they would accept the vaccination. 

\subsection{Vaccine certificate (green pass)}

The government has made another dubious decision on the wave of political influence.
From May 1, 2022 the green pass was eliminated and the obligation to use the ffp2 masks was maintained until June 15. It should be the opposite, with the obligation to wear masks only for anti-vaccine subjects. From 01/02/2021 to 10/01/2022 there were 46572 Covid positive deaths. Of these, 41227 deaths correspond to people who are not vaccinated and with an incomplete vaccination cycle (almost 90\% of all deaths), while 5345 deaths correspond to people vaccinated with a complete cycle (about 10\%). So, 90\% versus 10\%. The average age of the vaccinated subjects is 80 years and almost all had pre-existing diseases. If one wants to avoid deaths, one has to keep the green pass, because this protects the anti-vax population. Vaccinated people are already protected by the vaccine. The problem is that some people with other diseases have no place in hospitals. People were dying with about 150 deaths per day (May 13, 2022), as in a country of about 7-10 million unvaccinated people.
Again, the problem is avoiding contagion and not deaths.
In addition to scientific arguments, it was also about political-electoral and non-healthcare decisions. The above data belong to the ``Istituto Superiore di Sanit\`a"$\footnote{\url{https://www.epicentro.iss.it/coronavirus/sars-cov-2-decessi-italia#7}}$.

\subsection{Anti-vax healthcare workers} 

On November 1, 2022, the new Italian government allowed anti-vaccine doctors and healthcare workers to return to work, motivated, they say, by a worrying shortage of medical staff coupled with a decline in Covid-19 cases.
Anti-vaccine doctors were constantly spreading misinformation about the Covid vaccine
Levels of vaccine hesitancy among doctors are higher than expected, with 1 in 10 general practitioners not believing vaccines are the solution to the pandemic. The situation seems to be worse:
Among 625 doctors, 10\% disagree that vaccines are safe; 9.3\% disagree that vaccines are effective; and 8.3\% disagreed that they were important (Callaghan et al, 2022). Misconceptions about vaccines are a threat to global health, despite empirical evidence from the last 200 years confirming the effectiveness of vaccination (Gallegos et al., 2022). Unfortunately, pseudoscience is widespread. Most doctors who practice homeopathy (a pseudoscience) are anti-vaxers.
The failure to manage the pandemic is therefore evident, as shown in Figure 5, where
on May 4, 2021, Italy, the United States and Brazil were among the worst affected countries.

\section{Actionable recommendations}

A large number of COVID-19 infections can occur when an infected person enters a community. It is essential to simulate the infection (and death) process in advance, apply the appropriate control measures and mitigate the risk of spreading the virus. For this, an adequate pandemic plan is needed, managed by experts in statistics, differential equations and numerical analysis (Brauer, 2017).
One of the most used mathematical algorithms to describe the spread of an epidemic disease is the SEIR model, which we apply to calculate the number of people infected, recovered and dead based on the number of contacts, the probability of disease transmission, incubation and infection periods and mortality rate (Carcione et al., 2020). On the other hand, doctors and virologists must deal with hospital care and the vaccine.

With the vaccine available, it was mandatory to vaccinate the over 65s right from the start and this would have prevented tens of thousands of victims. Instead, the experts recommended vaccination for the risk of contagion (the so-called categories: school employees, lawyers, etc.) and not for the risk of death.

The deaths of thousands of people could have been avoided with a proper policy campaign.
Then, there is a competing human rights issue, ``liberty versus health" and ``economy versus health." Regarding the latter, Lesschaeve et al. (2021) conducted a survey with over 7000 respondents from South-Eastern Europe. The results show that, in general, public opinion was in favor of saving lives even at high economic costs. However, free-market views make people more accepting of higher casualties. 

A flexible pandemic plan should be developed through a multidisciplinary approach and structured around the following five levels:

(a) Epidemiological: Predictive modeling of different scenarios for the spread of the virus in the population, based on robust epidemiological surveillance data and evidence. Modern epidemiologists should adapt the more suitable existing models and develop new ones as needed.

(b) Medical: Research and development of treatments and vaccines to address the biological aspects of caring and preventing the disease. Healthcare workers must be trained to manage the pandemic at the national, regional level and locally, with simulations over time to ensure preparedness. Primary health care should be actively involved to guarantee system resilience.

(c) Logistics: Preparation of the healthcare response, including tests, protective equipment, hospital capacity, and vaccine administration. Engineers and economists should plan for needed resources and their territorial distribution. Coordination from the national government to local institutions with multiple decision-making centers (municipalities, provinces, regions), should be implemented to facilitate collective territorial coherence and minimize the possible consequences of a government crisis.

(d) Ethical: Balancing competing human rights, such as liberty, economy, and health. Sociologists, jurists and constitutionalists should discuss the appropriateness of the proposed policies.

(e) Communication: Design a permanent communication campaign using multimedia platforms to effectively convey information visually and verbally, such as through public service announcements, educational videos, infographics, and educational websites. Emphasize scientific dissemination on TV and other media instead talk shows. Present information in a clear, accessible, and engaging manner to reach a wide audience.

In addition, it is essential to vaccinate on the basis of the risk of death, and address the anti-vax problem with the appropriate information.
Furthermore, lack of staff and infrastructure in hospitals (e.g., doctors, healthcare workers, mechanical ventilators) due to underfunding of the healthcare system is another problem.
Moreover, healthcare workers must be specifically trained to manage the pandemic at the national, regional level and locally, with simulations at least once a year to be prepared for another pandemic.

\section{Conclusions}

It is crucial that we learn from our past mistakes and use the lessons learned to improve future responses to pandemics. In fact, we have identified several drawbacks in the management of the Italian pandemic: 1. The composition of the Scientific Committee; 2. Delays to prevent the first wave; 3. Failed prediction of the second wave; 4. Incorrect vaccination campaign; 5. Misuse of the vaccination certificate; 6. Media misinformation about vaccines and pandemic theory; 7. Contact tracing application failure; and 8. Persistence of the pandemic due to the anti-vaccination problem.

Freedom of expression and media reporting are essential for a democratic society, but adequate information on health risks is essential in an emergency. The spread of disinformation in any type of media has been as much of a threat to public health as the virus itself, for example, the spread and amplification of unverified stories, such as the spread of conspiracy theories, hoaxes, etc. This misinformation problem serves to undermine the pandemic response by eroding public trust and hampering efforts to control the spread of the virus and prevent anti-vaccine posturing. We should ask ourselves: in an emergency, are lockdowns and/or quarantines legal or a deprivation of liberty and limitation of human rights?
Is restricting the dissemination of inaccurate and false information during an emergency a deprivation of freedom of expression? If so, our means of dealing with a pandemic are very limited. Certainly it is a competing human rights issue, ``freedom versus health",
as it is also ``economy versus health" (or ``grow GDP or save lives?").

\vspace{0.5cm}

\section{Declarations}

{\bf List of abbreviations}: CTS: Scientific Technical Committee. WHO: World Health Organization. 
SEIR: Susceptible, Exposed, Infectious, Removed. 

\vspace{0.5cm}

{\bf Authors' contributions}: JMC: main idea and writing. JB: revising the draft and providing discussions.

\vspace{0.5cm}

{\bf Funding}: Not applicable.

\vspace{0.5cm}

{\bf Availability of data and materials}: Not applicable.

\vspace{0.5cm}

{\bf Conflict of interest}: The authors declare that they have no conflict of interest.

\vspace{0.5cm}

{\bf Ethics approval and consent to participate}: Not applicable.

\vspace{0.5cm}

{\bf Consent for publication}: JMC states that the content of the article expresses his personal view and does not represent the position of OGS in any way, also because the subject is unrelated to its institutional mission.

\vspace{0.5cm}

{\bf Competing interests}: The authors report there are no competing interests to declare


\newpage

\vspace{1cm}

\begin{figure}[htbp]
\hspace{-0.8cm} \includegraphics[scale=0.47]{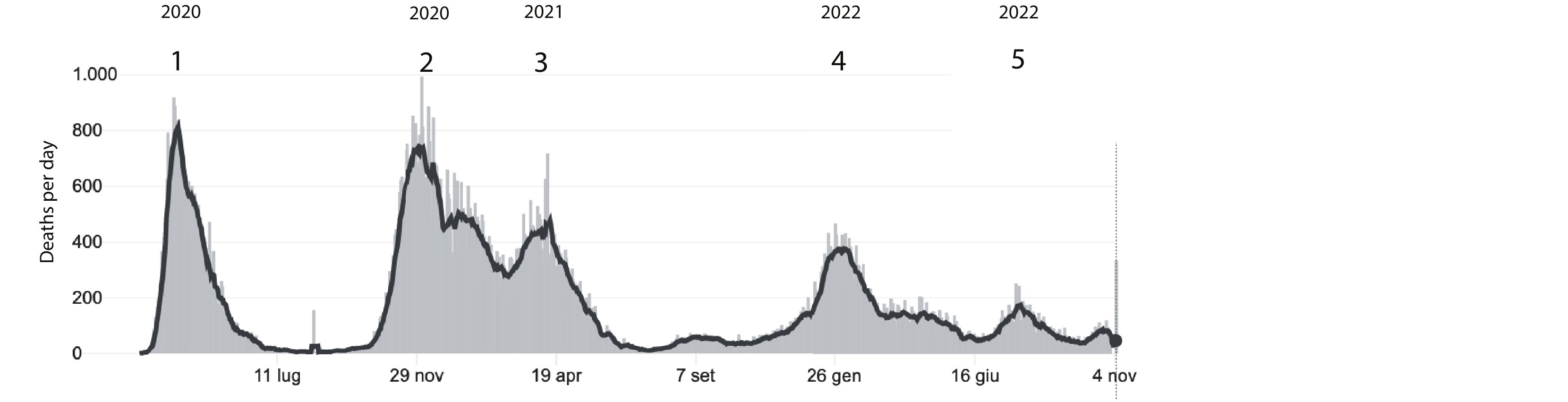}
\vspace{0cm}
\caption{
Five Italian waves of the Covid-19 pandemic. Deaths per day as a function of time.  Total deaths on November 7, 2022 are 179,000.
\blue https://github.com/CSSEGISandData/COVID-19 \black
}
\end{figure}

\begin{figure}[htbp]
\hspace{-5cm} \includegraphics[scale=0.6]{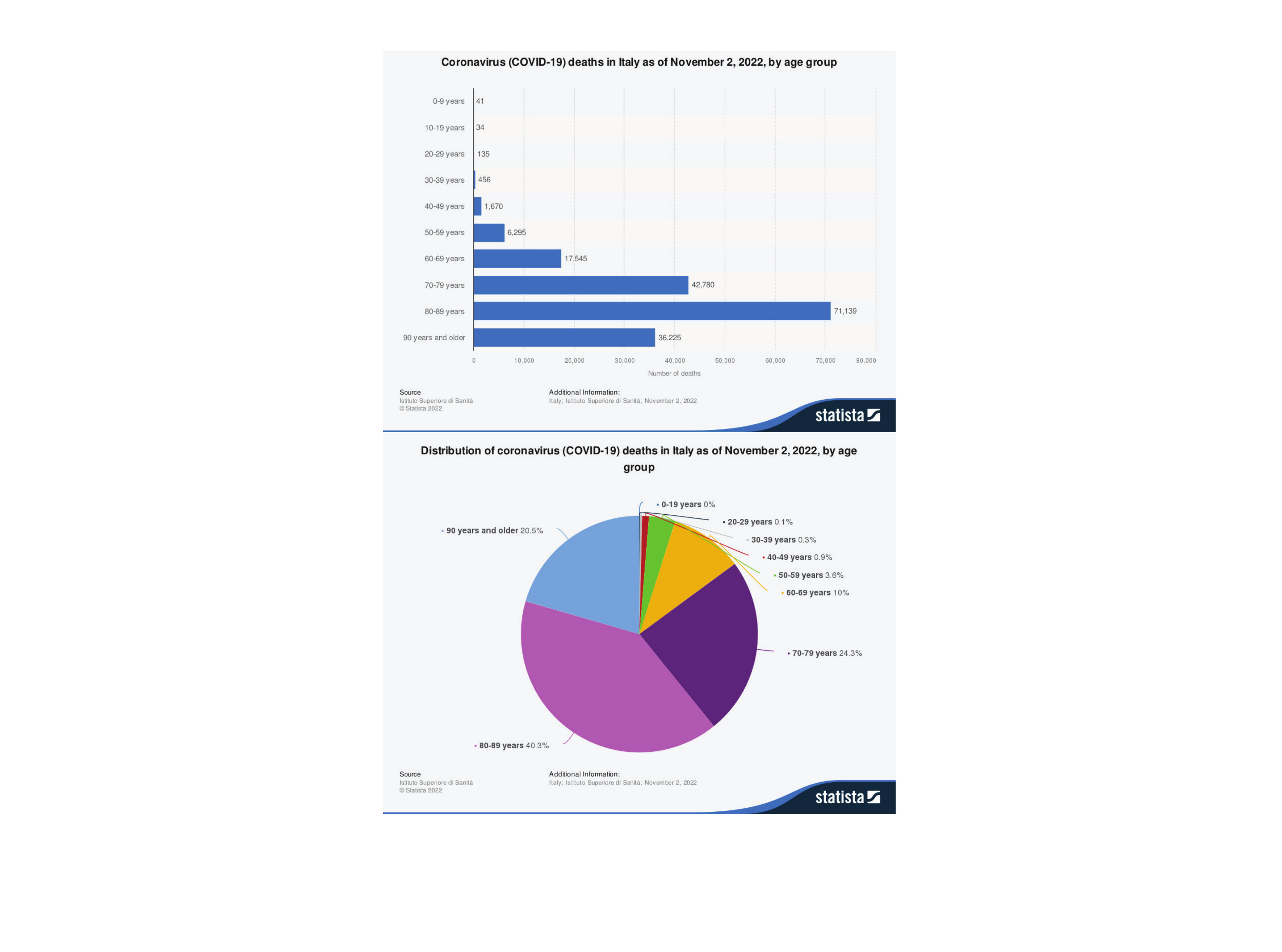}
\vspace{-2cm}
\caption{
Covid-19 deaths per age group in Italy on November 2, 2022. 
{\blue https://www.statista.com/statistics/1106367/coronavirus-deaths-distribution-by-age-group-italy/ \black}
}
\end{figure}

\begin{figure}[htbp]
\hspace{-15.5cm} \includegraphics[scale=0.5]{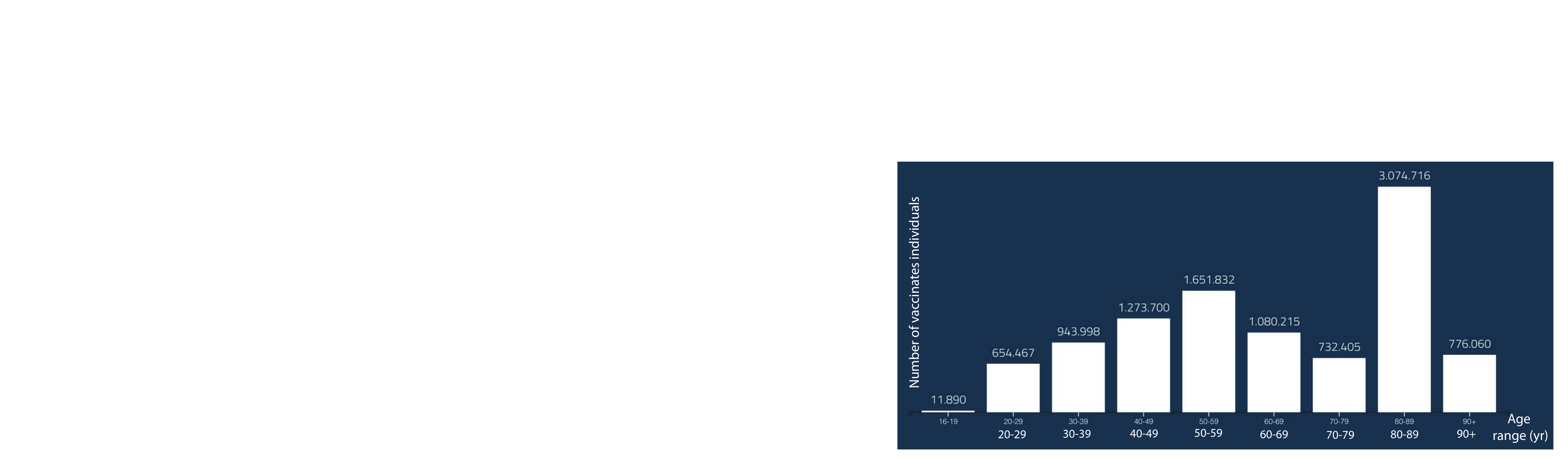}
\vspace{0cm}
\caption{
Vaccinated individuals in Italy on May 1, 2021, approximately 4 month after the
beginning of the vaccination campaign. \blue Data from https://github.com/pcm-dpc/COVID-19 \black  
}
\end{figure}

\begin{figure}[htbp]
\vspace{1cm}
\hspace{1cm} \includegraphics[scale=0.5]{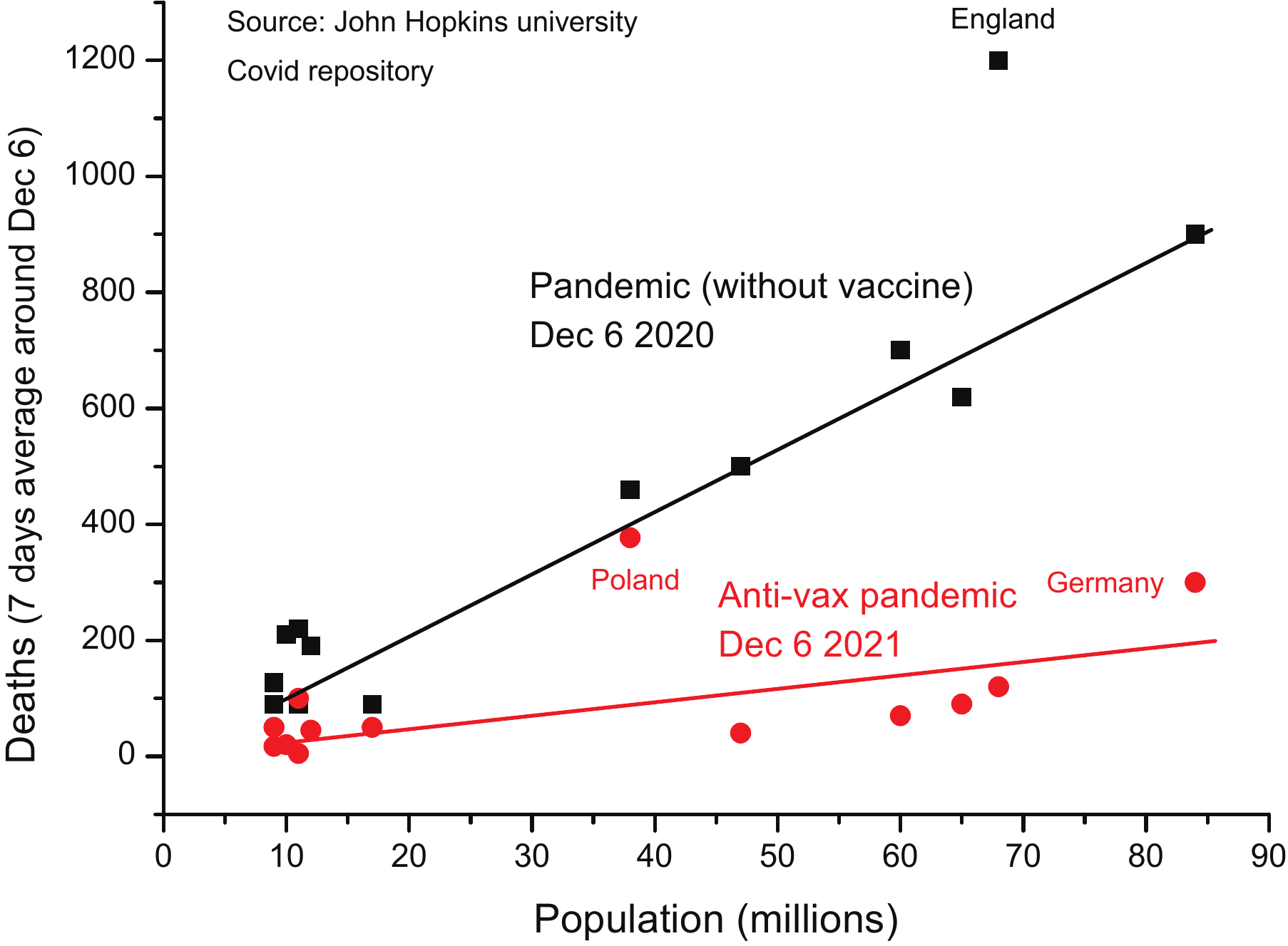}
\vspace{0cm}
\caption{
Number of people who died in an average of 7 days around 6 December 2020 (no vaccine present, black squares and line) and around 6 December 2021 (vaccine present, red dots and line) as a function of total population (in millions of people). 
Source: {\blue https://coronavirus.jhu.edu/map.html \black}
}
\end{figure}

\begin{figure}[htbp]
\vspace{1.5cm}
\hspace{2cm} \includegraphics[scale=0.45]{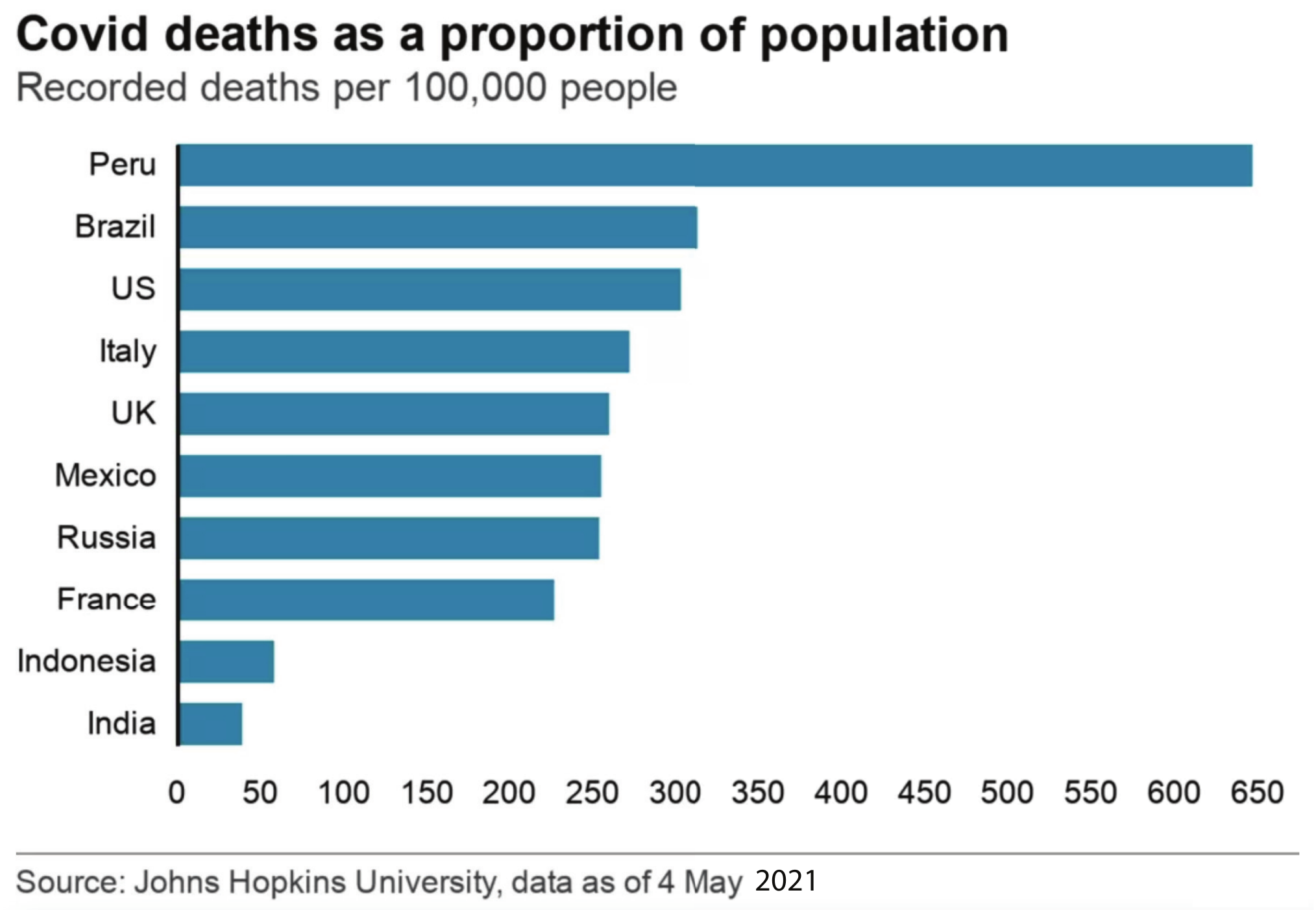}
\vspace{0cm}
\caption{
Covid-19 deaths as a proportion of the population (per 100,000 people) as of May 4, 2021. 
{ \blue https://www.bbc.com/news/61333847 \black }
}
\end{figure}

\end{document}